\documentstyle[aps,preprint,tighten,epsfig,prl]{revtex} % single-col/spacing
\def \invpb    {\relax{$\rm pb^{-1}$}}
\def \epem     {\relax{$e^+e^-$}}
\def \lplm     {\relax{\ell^{+}\ell^{-}}}
\def \lnu      {\relax{\ell^{\pm}\nu}}
\def \upum     {\relax{$\mu^+\mu^-$}}
\def \ttb      {\relax{$t\bar t$}}
\def \ccb      {\relax{$c\bar c$}}
\def \bbb      {\relax{$b\bar b$}}
\def \ppb      {\relax{p\bar p}}
\def \Wpm      {\relax{$W^{\pm}$}}
\def \Wp       {\relax{$W^{+}$}}
\def \Wm       {\relax{$W^{-}$}}
\def \Zz       {\relax{$Z^0$}}
\def \gluino   {\relax{\tilde{g}}}
\def \squark   {\relax{\tilde{q}}}
\def \slepton  {\relax{\widetilde \ell}}
\def \sneutrino {\relax{\widetilde \nu}}
\def \chichi   {\relax{${\tilde{\chi}_1^\pm \tilde{\chi}_2^0}$}}
\def \chizero  {\relax{\tilde{\chi}_1^0}}
\def \chione   {\relax{\tilde{\chi}_1^\pm}}
\def \chitwo   {\relax{\tilde{\chi}_2^0}}
\def \mumu     {\relax{\mu^+\mu^-}}
\def \ee       {\relax{e^+e^-}}
\def \ET       {\relax\ifmmode{E_{T}}\else{$E_{T}$}\fi}
\def \MET      {\relax{$\mbox{$\raisebox{.3ex}{$\not$}\ET$}$}}
\def \mgev     {GeV/$c^{2}$}
\def \sp       {\relax{$\;$}}
\def \dk       {\relax{\rightarrow}}
\def \gtsim    {\relax{\mathrel{\mathpalette\oversim >}}}
\def \ltsim    {\relax{\mathrel{\mathpalette\oversim <}}}
\def\oversim#1#2{\lower4pt\vbox{\baselineskip0pt \lineskip1.5pt
            \ialign{$\mathsurround=0pt#1\hfil##\hfil$\crcr#2\crcr\sim\crcr}}}
\font\eightit=cmti8
\def\r#1{\ignorespaces $^{#1}$}
\begin{document}
\draft
\title{
\begin{flushright}
FERMILAB-PUB-98/084-E\\[.2in]
\end{flushright}
Search for Chargino-Neutralino Associated Production
at the Fermilab Tevatron Collider}
\vspace{.2in}
\author {
\hfilneg
\begin{sloppypar}
\noindent
F.~Abe,\r {17} H.~Akimoto,\r {39}
A.~Akopian,\r {31} M.~G.~Albrow,\r 7 A.~Amadon,\r 5 S.~R.~Amendolia,\r {27} 
D.~Amidei,\r {20} J.~Antos,\r {33} S.~Aota,\r {37}
G.~Apollinari,\r {31}
% ...
\end{sloppypar}
\vskip .026in
\begin{center}
(CDF Collaboration)
\end{center}
\vskip .026in
\begin{center}
\r 1  {\eightit Argonne National Laboratory, Argonne, Illinois 60439} \\
\r 2  {\eightit Istituto Nazionale di Fisica Nucleare, University of Bologna,
I-40127 Bologna, Italy} \\
\r 3  {\eightit Brandeis University, Waltham, Massachusetts 02254} \\
\r 4  {\eightit University of California at Los Angeles, Los 
Angeles, California  90024} \\  
\r 5  {\eightit University of Chicago, Chicago, Illinois 60637} \\
\r 6  {\eightit Duke University, Durham, North Carolina  27708} \\
\r 7  {\eightit Fermi National Accelerator Laboratory, Batavia, Illinois 
60510} \\
\r 8  {\eightit University of Florida, Gainesville, FL  32611} \\
\r 9  {\eightit Laboratori Nazionali di Frascati, Istituto Nazionale di Fisica
               Nucleare, I-00044 Frascati, Italy} \\
\r {10} {\eightit University of Geneva, CH-1211 Geneva 4, Switzerland} \\
\r {11} {\eightit Harvard University, Cambridge, Massachusetts 02138} \\
\r {12} {\eightit Hiroshima University, Higashi-Hiroshima 724, Japan} \\
\r {13} {\eightit University of Illinois, Urbana, Illinois 61801} \\
\r {14} {\eightit Institute of Particle Physics, McGill University, Montreal 
H3A 2T8, and University of Toronto,\\ Toronto M5S 1A7, Canada} \\
\r {15} {\eightit The Johns Hopkins University, Baltimore, Maryland 21218} \\
\r {16} {\eightit Institut f\"{u}r Experimentelle Kernphysik, 
Universit\"{a}t Karlsruhe, 76128 Karlsruhe, Germany} \\
\r {17} {\eightit National Laboratory for High Energy Physics (KEK), Tsukuba, 
Ibaraki 305, Japan} \\
\r {18} {\eightit Ernest Orlando Lawrence Berkeley National Laboratory, 
Berkeley, California 94720} \\
\r {19} {\eightit Massachusetts Institute of Technology, Cambridge,
Massachusetts  02139} \\   
\r {20} {\eightit University of Michigan, Ann Arbor, Michigan 48109} \\
\r {21} {\eightit Michigan State University, East Lansing, Michigan  48824} \\
\r {22} {\eightit University of New Mexico, Albuquerque, New Mexico 87131} \\
\r {23} {\eightit The Ohio State University, Columbus, OH 43210} \\
\r {24} {\eightit Osaka City University, Osaka 588, Japan} \\
\r {25} {\eightit Universita di Padova, Istituto Nazionale di Fisica 
          Nucleare, Sezione di Padova, I-35131 Padova, Italy} \\
\r {26} {\eightit University of Pennsylvania, Philadelphia, 
        Pennsylvania 19104} \\   
\r {27} {\eightit Istituto Nazionale di Fisica Nucleare, University and Scuola
               Normale Superiore of Pisa, I-56100 Pisa, Italy} \\
\r {28} {\eightit University of Pittsburgh, Pittsburgh, Pennsylvania 15260} \\
\r {29} {\eightit Purdue University, West Lafayette, Indiana 47907} \\
\r {30} {\eightit University of Rochester, Rochester, New York 14627} \\
\r {31} {\eightit Rockefeller University, New York, New York 10021} \\
\r {32} {\eightit Rutgers University, Piscataway, New Jersey 08855} \\
\r {33} {\eightit Academia Sinica, Taipei, Taiwan 11530, Republic of China} \\
\r {34} {\eightit Texas A\&M University, College Station, Texas 77843} \\
\r {35} {\eightit Texas Tech University, Lubbock, Texas 79409} \\
\r {36} {\eightit Istituto Nazionale di Fisica Nucleare, University of Trieste/
Udine, Italy} \\
\r {37} {\eightit University of Tsukuba, Tsukuba, Ibaraki 315, Japan} \\
\r {38} {\eightit Tufts University, Medford, Massachusetts 02155} \\
\r {39} {\eightit Waseda University, Tokyo 169, Japan} \\
\r {40} {\eightit University of Wisconsin, Madison, Wisconsin 53706} \\
\r {41} {\eightit Yale University, New Haven, Connecticut 06520} \\
\end{center}
\date{\today}
}

\maketitle
\begin{abstract}

We have searched in $p \overline{p}$ collisions at $\sqrt{s}$ = 1.8~TeV for
events with three charged leptons and missing transverse energy.  In the
Minimal Supersymmetric Standard Model, we expect trilepton events from
chargino-neutralino ($\chione \chitwo$) pair production, with subsequent decay
into leptons.  We observe no candidate  $e^+e^-e^\pm$, $e^+e^-\mu^\pm$,
$e^\pm\mu^+\mu^-$ or $\mu^+\mu^-\mu^\pm$ events  in 106 pb$^{-1}$ integrated
luminosity.  We present limits on the sum of the branching ratios times cross
section for the four channels: $\sigma_{\chione\chitwo}\cdot
$BR$(\chione\chitwo\rightarrow 3\ell+X) < 0.34$~pb, M$_\chione > $ 81.5
\mgev\sp and M$_\chitwo > $ 82.2 \mgev\sp for $\tan\beta=2$, $\mu
=-600$~\mgev\sp and M$_\squark=$ M$_\gluino$. 

\end{abstract}
\vspace{.2in}
\pacs{PACS numbers: 14.80.Ly, 12.60.Jv, 13.85.Rm}
\vspace{.2in}

\narrowtext

The Minimal Supersymmetric Standard Model (MSSM)~\cite{basicMSSM} contains two
Higgs doublets and supersymmetric partners to all the Standard Model (SM)
particles. The superpartners of the electroweak gauge bosons and Higgs bosons
are two charged and four neutral fermions ($\widetilde{\chi}$'s). Further
assumptions, namely the Grand Unified Theory hypothesis provided by
Supergravity~\cite{basicSUGRA}, Supergravity-inspired slepton/sneutrino mass
constraints~\cite{rge}, degeneracy of five of the squarks, and R-parity
conservation, lead to models with only six parameters. R-parity conservation
implies the creation of superpartners in pairs and the stability of the
lightest supersymmetric partner (LSP). Within this framework we expect, for
certain regions of parameter space, a measurable rate for the reaction
$q\bar{q}'\dk\chione\chitwo$, where $\chione\dk\chizero\lnu$ and
$\chitwo\dk\chizero\lplm$, and $\chizero$ is the LSP. The $\nu$ and two LSPs do
not interact with the detector and manifest themselves as missing energy. The
resulting final state is three isolated charged leptons plus missing
energy~\cite{Trilepton}. In this Letter, we report on a search for direct
production of $\chione\chitwo$, via virtual \Wpm\sp $s$-channel and virtual
squark $t$-channel diagrams, in the trilepton channels $e^+e^-e$, $e^+e^-\mu$,
$e\mu^+\mu^-$ and $\mu^+\mu^-\mu$. Additional trilepton production arising from
squark and gluino cascade decays was not included. We add 87 \invpb\sp of data
recorded in 1994-95 to a previously analyzed sample of 19 \invpb\sp collected
in 1992-93~\cite{run1a_paper}.

The Collider Detector at Fermilab (CDF) is described in detail
elsewhere~\cite{CDFdet}. The components of the detector relevant to this
analysis are the vertex chamber, which provides $r$-$z$ tracking information;
the central tracking chamber, which is situated inside a 1.4 T solenoidal
magnetic field and provides a combination of $r$-$\phi$, $z$ and transverse
momentum ($p_T$) information for charged particles; the central
($\vert\eta\vert < 1.1$) and endplug ($1.1 < |\eta| < 2.4$) electromagnetic
calorimeters, which are located outside the solenoid and are segmented in a
projective tower geometry; and the central muon chambers. We define 
pseudorapidity $\eta \equiv -\ln\tan(\theta/2)$ and $\theta$ and $\phi$ to be
the polar and azimuthal angles with respect to the beam axis. 

We begin with a sample of 87 pb$^{-1}$ recorded in 1994-95, which contains $3.3
\times 10^6$ events that have an electron or muon with $p_T > 8$ GeV/c and
$|\eta^e|< 1.1$ or $|\eta^\mu|< 0.6$, and an additional charged lepton with
$p_T > 3$ GeV/c and $|\eta^{e}| < 2.4$ or $|\eta^{\mu}| < 1.0$. We select
events by requiring an electron with $E_T^e > 11$ GeV and $|\eta^e|< 1.1$ or a
muon with $p_T^\mu > 11$ GeV/c and $|\eta^\mu|< 0.6$. We require two additional
charged leptons with $E_T^e > 5$ GeV and $|\eta^e|< 2.4$ or $p_T^\mu > 4$ GeV/c
and $|\eta^\mu|< 1.0$. At least one lepton passing the high threshold cut must
pass stringent lepton identification
cuts~\cite{run1a_paper,CDFdet,top,wmass,mythesis}. To improve the integrity of
these events, we require that all three leptons originate from a common vertex
within 60 cm of the center of the detector. The 60 cm requirement is to
preserve the projective tower geometry of the calorimetry. We find 59 events
meeting these requirements.

The principal backgrounds to the $\chione\chitwo$\sp signature are real
trilepton events from \Wpm\Zz, \Zz\Zz, \ttb\sp and \bbb\sp and dilepton plus
fake lepton\cite{fakes} events from \Wp\Wm, \Zz\sp and the Drell-Yan process. 
To remove background from \bbb, \ccb\sp and \ttb\sp production and fake
leptons, each lepton must be isolated, where isolation is defined by requiring
less than 2 GeV $E_T$ in the calorimeter inside an $\eta$-$\phi$ cone of radius
$\Delta R \equiv \sqrt{(\Delta\phi)^2 + (\Delta\eta)^2}$ = 0.4 around the
lepton, excluding the energy deposited by the lepton. There must be at least
one \epem\sp or \upum\sp pair, the $\eta$-$\phi$ distance between any two
leptons ($\Delta R_{\ell \ell}$) must be greater than 0.4 (to remove background
from \bbb\sp production, as well as some anomalously reconstructed cosmic ray
events) and the difference in azimuthal angle between the two highest $p_T$
leptons ($\Delta \phi_{\ell_1 \ell_2}$) in the event must be less than
170$^\circ$ (to remove background from the Drell-Yan process and cosmic
rays)~\cite{mythesis}. Events containing a same flavor $\lplm$\sp pair with
invariant mass in the regions of the resonances $J/\psi$ (2.9-3.3~\mgev),
$\Upsilon$ (9-11~\mgev) and \Zz\sp (75-105~\mgev) are removed. These
requirements select 6 events (see Table~\ref{events_left}). In the previous
data sample~\cite{run1a_paper} these criteria selected zero events.

The presence of two LSPs and a neutrino in the final state of the signal can
lead to substantial missing transverse energy (\MET). The dominant remaining
backgrounds, \bbb\sp production and the Drell-Yan process, do not have
significant \MET.  As seen in Table~\ref{events_left}, requiring \MET $>$ 15
GeV reduces the background by 85\% while retaining 82\% of the expected signal
for M$_\chione\approx$ 70 \mgev. No events pass the \MET\sp cut.

For the remainder of the analysis we combine this data with the previous
19$\pm$1~\invpb\sp sample~\cite{run1a_paper} for a total Run I integrated
luminosity ($\int{\cal L}dt$) of 106$\pm$7 \invpb\sp and zero candidate
trilepton events.

To determine the SM background and the signal acceptance, we use the ISAJET
Monte Carlo program~\cite{ISAJET_overview} with a CDF detector simulation. For
background due to vector boson pair production we use theoretical calculations
of cross sections and branching ratios~\cite{dibosons}. For background due to
\ttb\sp production and the Drell-Yan process we use cross sections measured by
CDF~\cite{top}. The rate of lepton misidentification was determined from a \Wpm
$\rightarrow\ell\nu$ data sample to be (0.29$\pm$0.04)\% per event. After all
cuts are applied the total expected background is 1.2$\pm$0.2 events in
106~pb$^{-1}$. 

The total detection efficiency ($\epsilon^{tot}$) is a product of the trigger
efficiency, the isolation efficiency, the lepton identification efficiency and
a geometric and kinematic acceptance factor. The triggers used were single
lepton and dilepton triggers, with efficiencies of $\epsilon_e^{trig} =
(87^{+4}_{-5})\%$ above 11 GeV and $\epsilon_\mu^{trig} = (87\pm3)\%$ above 11
GeV/c. We determine the lepton isolation and identification efficiencies by
studying the second lepton in \Zz$\dk\lplm$ events. The isolation efficiency is
(90$\pm$4)\% per lepton. The per-event trilepton identification efficiency
ranges from (59$\pm$1)\% to (82$\pm$1)\%, depending on the combination of
leptons in the event. The geometric and kinematic acceptance is determined
using ISAJET and the CDF detector simulation.  The total efficiency
($\epsilon^{tot}$) is mainly a function of the $\chione$\sp and $\chitwo$\sp
masses, which are nearly equal for the region of the search. The efficiency
increases linearly from 3\% at 50 GeV/c$^2$ to 12\% at 100 GeV/c$^2$, because
massive $\chione$\sp and $\chitwo$\sp lead to more central and more energetic
leptons which are detected with higher efficiency. 

We see no signal candidates and thus set limits on the available parameter
space. A particular point in parameter space is excluded if the predicted
number of events exceeds the number of events ($s$) expected at the 95\%
confidence level limit given that zero events were observed. The predicted
number of events is a function of the cross section times branching ratio
($\sigma(\ppb \dk \chione\chitwo + X)\cdot $BR$(\chione\chitwo \rightarrow
3\ell+X)$) and $\epsilon^{tot}\cdot\int{\cal L}dt$. We calculate cross section
times branching ratio ($\sigma\cdot$ BR) using ISAJET 7.20 with
CTEQ-3L~\cite{cteq} parton distribution functions and calculate $s$ by
convolving the total systematic uncertainty as a Gaussian smearing with a
Poisson distribution. The total systematic uncertainty is 15\%, which includes
uncertainty in the total integrated luminosity ($\pm7\%$), the parton
distribution ($\pm7\%$), the trigger efficiency ($\pm6\%$), and the
trilepton-finding efficiency ($\pm2\%$), leading to $s=3.1$. To calculate the
uncertainty due to the parton distribution function we compare CTEQ-3L with a
variety of other parton distribution functions. We use the largest uncertainty
in the efficiency of any single lepton trigger for all events.

Using the model assumptions listed in the first paragraph, four parameters
determine the $\chione$\sp and $\chitwo$\sp masses, production cross sections
and decay branching ratios: the ratio of the Higgs vacuum expectation values
($\tan\beta$), the Higgsino mass parameter ($\mu$), the gluino mass
(M$_\gluino$) and the squark-to-gluino mass ratio (M$_\squark/$M$_\gluino$). To
make the analysis more independent of details of the Higgs sector, we
consider a region in the MSSM where there is no significant chargino or
neutralino branching fraction into Higgs particles. Technically, we do this by
choosing the mass of the pseudoscalar Higgs (M$_A$) to be above the
chargino/neutralino mass and use M$_A$ = 500 \mgev. The production and decay of
$\chione\chitwo$\sp are independent of the remaining MSSM parameter, the
trilinear top squark coupling ($A_t$). We fix $A_t = \mu/\tan\beta$ for
consistency with other CDF analyses~\cite{stephan}. The search is more
sensitive at low values of $\tan\beta$; $\tan\beta \gtsim 10$ leads to higher
branching ratios to $\tau$'s, for which we do not search. We consider
$1.1\le\tan\beta\le8$.  We use $-1000$ GeV/c$^2 <\mu<-200$ GeV/c$^2$ because
the search is more sensitive to negative values of $\mu$ and $|\mu|$ is
expected to be on the order of the energy scale at which supersymmetric
phenomena should be observable. Also, small $|\mu|$ increases the Higgsino
content of the $\chione$\sp and $\chitwo$, which decreases the branching ratio
to leptons. ISAJET requires M$_\gluino$ and M$_\squark$ as input parameters to
calculate M$_\chione$. The  slepton/sneutrino mass constraints~\cite{rge} use
M$_\gluino$ and M$_\squark$ to determine M$_\slepton$ and M$_\sneutrino$; large
differences in M$_\gluino$ and M$_\squark$ lead to heavy sleptons and
sneutrinos and decreases the branching ratio to leptons. Thus, this analysis
considers M$_\squark/$M$_\gluino>1$ to avoid invisible decays through light
sneutrinos and M$_\squark/$M$_\gluino<2$ to enhance leptonic final states. For
the regions of parameter space we examine M$_\gluino\approx3$M$_{\chione}$, so
we use 150 \mgev$\le$ M$_\gluino\le$ 340 \mgev.

\begin{figure}
\center\mbox{\epsfig{file=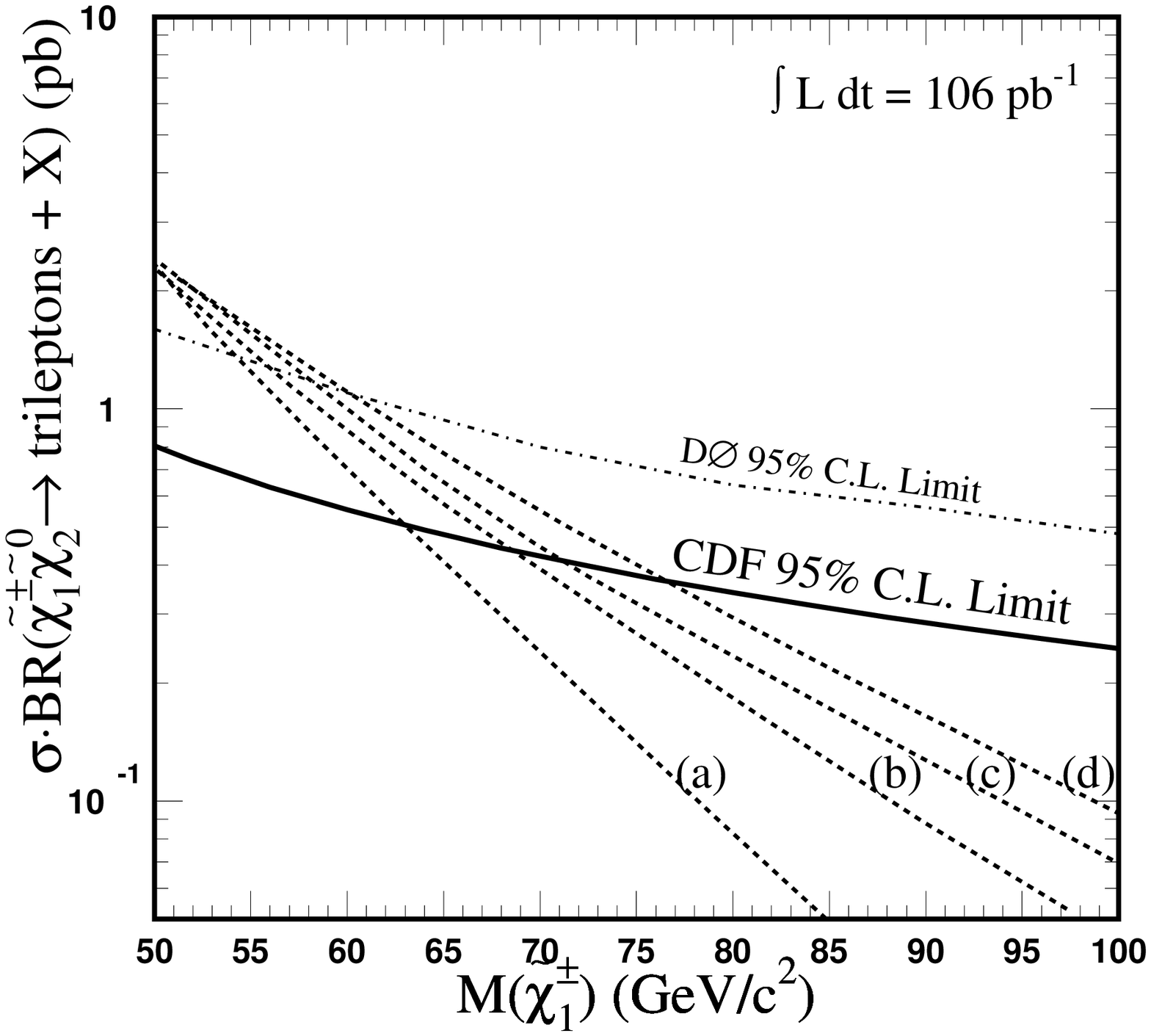,width=3.4in}}
\caption{ $\sigma\cdot $BR($\chione\chitwo\dk3\ell + X)$ versus $\chione$\sp
mass for  representative points in the MSSM parameter space, namely $\mu=-400$
\mgev, tan $\beta$ = 2 and (a) M$_\squark$/M$_\gluino$ = 2.0, (b)
M$_\squark$/M$_\gluino$ = 1.5, (c) M$_\squark$/M$_\gluino$ = 1.2 and (d)
M$_\squark$/M$_\gluino$ = 1.0. BR is the summed branching ratio into the four
trilepton modes ($e^+e^-e$, $e^+e^-\mu$, $e\mu^+\mu^-$ and $\mu^+\mu^-\mu$).
The solid line is the 95\% confidence level upper limit based on an observation
of zero events. We set a mass limit of 77.0~\mgev\sp when
M$_\squark$=M$_\gluino$. All MSSM points in this plot yield three body decays
of the $\chione$\sp and $\chitwo$. The D\O\ limit~[15] is for single
trilepton mode which we scale up by 4 to match our notation.}
\label{limit}
\end{figure}

\begin{figure}
\center\mbox{\epsfig{file=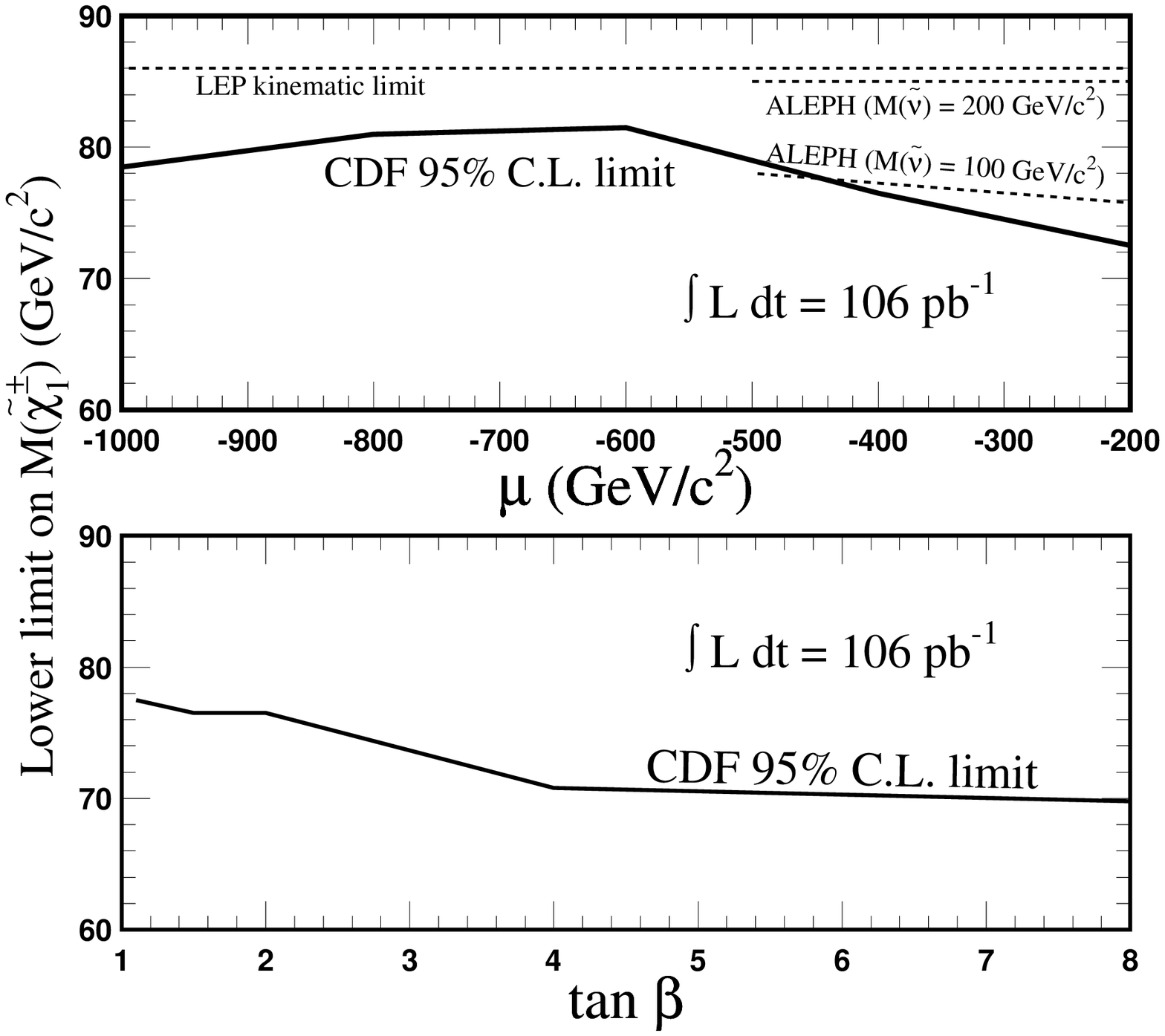,width=3.4in}}
\caption{ The experimental limit on M$_{\chione}$ as a function of $\mu$ for
$\tan\beta=2$ and M$_\squark$ = M$_\gluino$ (upper) and as a function of
$\tan\beta$ for $\mu=-400$ \mgev\sp and M$_\squark$ = M$_\gluino$ (lower). The
ALEPH limits  shown~[16] are from a search for all possible final
states. In this analysis M$_\sneutrino\approx$~100~GeV/c$^2$. Minimal SUGRA
models favor the region of small $|\mu|$ values. }
\label{mu}
\end{figure}

Figure~\ref{limit} shows the limit for a few representative points in the
M$_{\chione}-(\sigma\cdot$BR) plane. All points above the solid line are
excluded. For comparison, we include the result of the D\O\
collaboration~\cite{d0-limit}. D\O\ reports the {\it average} $\sigma\cdot$BR;
we use the sum. Figure~\ref{mu} shows the limit on M$_{\chione}$ as a function
of $\mu$ and $\tan\beta$. These limits are compared to the limits from
ALEPH~\cite{aleph} in Figure~\ref{mu}. The ALEPH result is from a search for
all possible final states. The OPAL, L3 and DELPHI collaborations report
similar limits~\cite{opal}.

\begin{figure}
\center\mbox{\epsfig{file=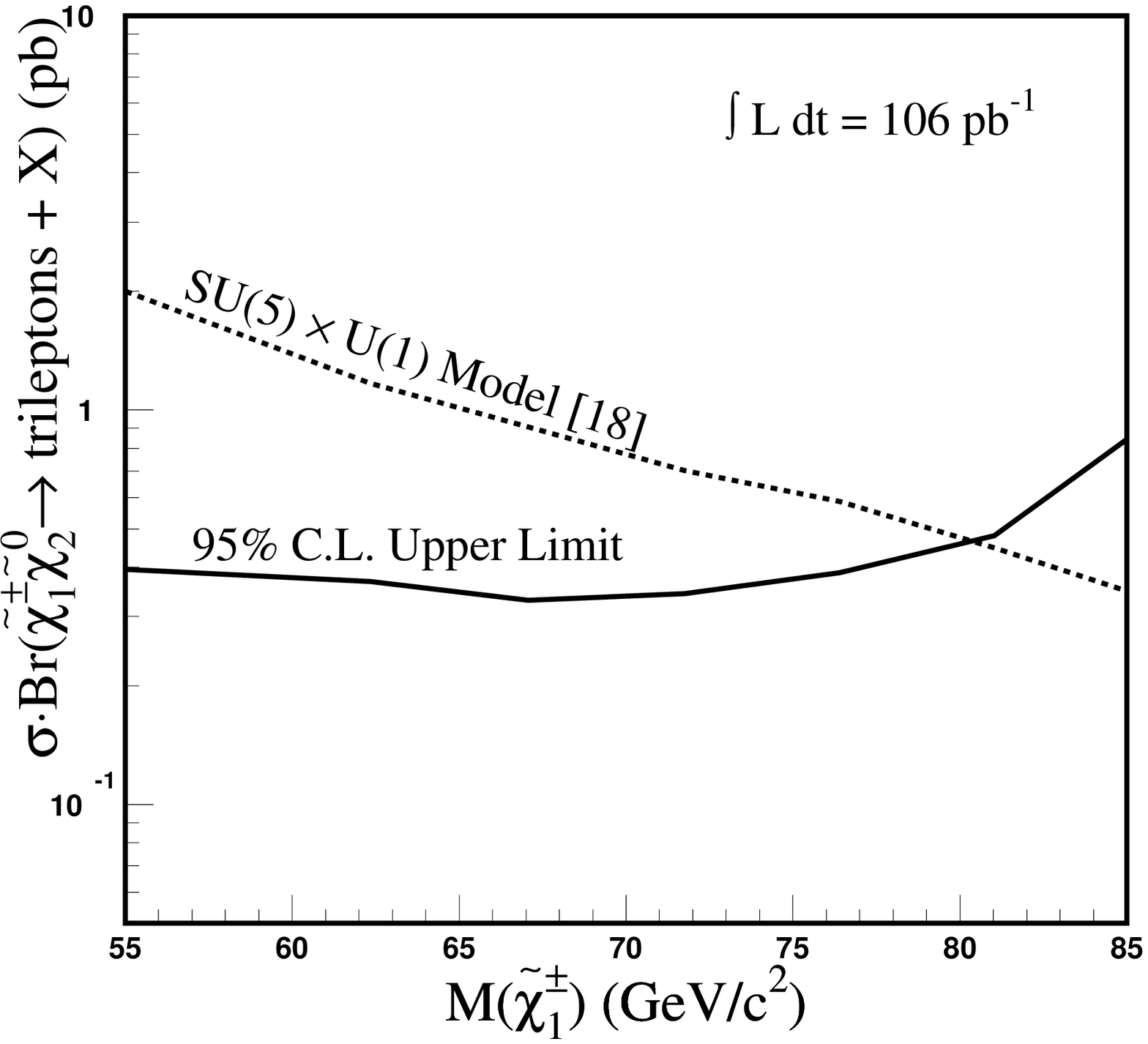,width=3.4in}}
\caption{$\sigma\cdot $BR($\chione\chitwo\dk3\ell+X)$ versus $\chione$\sp mass
for the SU(5) $\times$ U(1) model~[18]. This sets a mass limit of
80.5~\mgev. In this model, the $\slepton_R$ is lighter than the $\chitwo$\sp,
resulting in two body decays of the $\chitwo$. Note that the acceptance for
events from this model decreases at large $\chione$\sp mass. In this region, the
LSP mass approaches that of the $\slepton$, resulting in soft final state
leptons which are difficult to detect.}
\label{stringy}
\end{figure}

We also examined a string-inspired SU(5) $\times$ U(1) one-parameter
supergravity model~\cite{su5_prd}. This model requires
M$_{\chione}\ltsim$ 87\mgev\sp and M$_\chitwo\ltsim$ 91\mgev\sp and has a
nearly maximized trilepton branching ratio via
$\chitwo\rightarrow\slepton_R\ell$ and $\slepton_R\rightarrow\ell\chizero$.
As shown in Figure~\ref{stringy}, we exclude most of this model and set
M$_{\chione}>80.5$~GeV/c$^2$, M$_\chitwo >$~86.7~\mgev\sp and  $\sigma\cdot
$BR$(\chione\chitwo\dk 3\ell+X) < 0.48$~pb.

In conclusion, we find no evidence for \chichi\sp production in 1.8 TeV
$\ppb$\sp collisions and set limits on $\chione$\sp and $\chitwo$\sp masses and
$\sigma\cdot $BR within the framework of MSSM models which have
M$_{\chione}\approx $M$_{\chitwo}\approx$2M$_\chizero$ and three-body decays of
$\chione$\sp and $\chitwo$. The strongest limit is $\sigma_{\chione\chitwo}\cdot
$BR$(\chione\chitwo\rightarrow$3$\ell$+$X)<$~0.34~pb, M$_{\chione}
>$~81.5~\mgev\sp and M$_\chitwo >$~82.2~\mgev\sp for $\tan\beta=2$, $\mu =
-600$~\mgev\sp and M$_\squark=$ M$_\gluino$.

We thank the Fermilab staff and the technical staffs of the participating
institutions for their vital contributions. This work was supported by
the U.S. Department of Energy and National Science Foundation; the Italian
Istituto Nazionale di Fisica Nucleare; the Ministry of Education, Science and
Culture of Japan; the Natural Sciences and Engineering Research Council of
Canada; the National Science Council of the Republic of China; the A. P. Sloan
Foundation.

\begin{table}
\caption{ Events remaining after each cut in the 1994-95 data (87 pb$^{-1}$).
One \Zz\sp event and one $J/\psi$ event are removed with the resonance cuts.
For comparison we indicate the expected background (BG) and an expected signal
from a representative MSSM Monte Carlo (MC) sample (M$_\squark$ =
M$_\gluino=200$ \mgev, $\tan\beta=2$, $\mu$ = $-$400 \mgev,
M$_{\chione}\simeq$ M$_\chitwo\simeq 70$ \mgev,
$\sigma_{\chione\chitwo}=4.8$~pb, $\epsilon^{tot} = 6.7\%$).}
\begin{center}
\begin{tabular}{lrrr}
&&\multicolumn{1}{c}{Expected}&\multicolumn{1}{c}{MSSM}\\
\multicolumn{1}{c}{Cut}&\multicolumn{1}{c}{Events}
&\multicolumn{1}{c}{BG}&\multicolumn{1}{c}{MC}\\ \hline
Dilepton data set                          &3,270,488&&\\ 
Trilepton data set                         &59       &&\\ 
$\circ$ Isolation $<$ 2 GeV                &23       &&\\ 
$\circ$ Require $\ee$ or $\mumu$           &23       &&\\ 
$\circ$ $\Delta  R_{\ell\ell} > 0.4$       & 9       &&\\ 
$\circ$ $\Delta\phi_{\ell_1\ell_2} < 170^\circ$&8    &9.6$\pm$1.5&6.2$\pm$0.6\\ 
$\circ$ $J/\psi$, $\Upsilon$, \Zz\sp removal  & 6    &6.6$\pm$1.1&5.5$\pm$0.5\\ 
$\circ$ \MET$> 15$ GeV                     & 0       &1.0$\pm$0.2&4.5$\pm$0.4\\ 
\hline
Total Run I data (106 \invpb)              & 0       &1.2$\pm$0.2&5.5$\pm$0.4\\
\end{tabular}
\end{center}
\label{events_left}
\end{table}

\begin{thebibliography}{99}

\bibitem{basicMSSM} H.P.~Nilles, Phys. Rep. {\bf 110}, 1 (1984);  H.E.~Haber
	and G.L.~Kane, Phys. Rep. {\bf 117}, 75 (1985).

\bibitem{basicSUGRA}  A.H.~Chamseddine {\it et al.},  Phys. Rev. Lett. {\bf
	49}, 970 (1982); R.~Barbieri {\it et al.}, Phys. Lett. {\bf B119}, 343
	(1982); L.~Hall {\it et al.}, Phys. Rev. {\bf D27}, 2359 (1983). For a
	review, see R.~Arnowitt and P.~Nath,  ``Supersymmetry and
	Supergravity,'' VII$^{th}$ Swieca Summer School,  Campos de Jordao,
	Brazil, (World Scientific, Singapore, 1994).

\bibitem{rge}  L.E.~Iba$\tilde{{\rm n}}$ez {\it et al.},  Nucl. Phys. {\bf
	B256}, 218 (1985); G.G.~Ross and R.G.~Roberts, Nucl. Phys. {\bf B377},
	571 (1992); R.~Arnowitt and P.~Nath, Phys. Rev. Lett. {\bf 69}, 725
	(1992); S.~Kelley {\it et al.}, Nucl. Phys. {\bf B398}, 31 (1993);
	G.L.~Kane {\it et al.}, Phys. Rev. {\bf D49}, 6173 (1994). In this
	analysis we use formulae in H.~Baer {\it et al}., Phys. Rev. {\bf 
	D47}, 2739 (1993). 

\bibitem{Trilepton} P.~Nath and R.~Arnowitt,  Mod. Phys. Lett. {\bf A2}, 331
	(1987); R.~Barbieri {\it et al.}, Nucl. Phys. {\bf B367}, 28 (1991);
	J.L.~Lopez {\it et al.}, Phys. Rev. {\bf D48}, 2062 (1993); H.~Baer and
	X.~Tata, Phys. Rev. {\bf D48}, 5175 (1993).

\bibitem{run1a_paper} F.~Abe {\it et al.} (CDF Collaboration), Phys. Rev. Lett.
{\bf 76}, 4307 (1996).

\bibitem{CDFdet} F.~Abe {\it et al.} (CDF Collaboration), Nucl. Instrum.
	Methods Phys. Res., Sect. A {\bf 271}, 387 (1988); F.~Abe {\it et al.},
	Phys. Rev. {\bf D50}, 2966 (1994).

\bibitem{top} F.~Abe {\it et al.} (CDF Collaboration), Phys. Rev. Lett. {\bf
        74}, 2626 (1994); F.~Abe {\it et al.} (CDF Collaboration), Phys. Rev. 
	{\bf D49}, 1 (1994).

\bibitem{wmass} F. Abe {\it et al.} (CDF Collaboration), Phys. Rev. {\bf D52},
         4784 (1995).

\bibitem{mythesis} B. Tannenbaum, Ph. D. dissertation, University of New Mexico,
         NMCPP 97/12 (1997).

\bibitem{fakes} ``Fake lepton'' includes both non-prompt leptons such as
        decay-in-flight muons and non-leptonic objects passing the lepton
        identification cuts.

\bibitem{ISAJET_overview}  H.~Baer {\it et al.}, ``Simulating Supersymmetry
	with ISA\-JET 7.0/ISASUSY 1.0,''  Proc. of  Workshop on Physics at
	Current Accelerators and  the Supercollider,  (Argonne Nat. Lab.,
	1993). We use ISAJET V7.20.

\bibitem{dibosons} J. Ohnemus, Phys. Rev. {\bf D 44}, 1403 (1991) and Phys.
        Rev. {\bf D44}, 3477 (1991); J. Ohnemus and J.F. Owens, Phys. Rev. 
	{\bf D 43}, 3626 (1991).

\bibitem{cteq} H.L. Lai {\it et al.} (CTEQ Collaboration), Phys. Rev. {\bf
         D51}, 4763 (1995).

\bibitem{stephan} F.~Abe {\it et al.} (CDF Collaboration), Phys. Rev. Lett.
        {\bf76}, 2006 (1996); F.~Abe {\it et al.} (CDF Collaboration), Phys. 
	Rev. {\bf D56}, 1357 (1997).

\bibitem{d0-limit} B. Abbott {\em et al.} (D\O\sp Collaboration), Phys. Rev.
        Lett. {\bf80}, 1591 (1998).

\bibitem{aleph} R. Barate {\it et al.} (ALEPH Collaboration), CERN-PPE-97-128
        (1997).

\bibitem{opal}  K. Ackerstaff {\it et al.} (OPAL Collaboration),
        CERN-PPE-97-083 (1997); M. Acciarri {\it et al.} (L3 Collaboration), 
        CERN-PPE/97-130 (1997); P.Abreu {\it et al.} (DELPHI Collaboration),
        CERN-PPE/97-107 (1997). 

\bibitem{su5_prd} J. Lopez {\it et al.}, Phys. Rev. {\bf D52},
        4178  (1995) and Phys. Rev. {\bf D53}, 5253 (1996).

\end{thebibliography}
\end{document}